# Concentration regimes in salt-free aqueous xanthan solutions under shear

Ammar El Menayyir, Markus Neuner[1], Polina Fuks, Vahid A. Z. Alashloo, Halim Altuntas, Zehau Luo, Melike Özgül, Claudia Seeberger, Sharadwata Pan, Andreas Wierschem

Institute of Fluid Mechanics, Friedrich-Alexander-Universität Erlangen-Nürnberg (FAU), Cauerstr. 4, 91058 Erlangen, Germany

[1] Current address: NETZSCH-Gerätebau GmbH, Wittelsbacherstr. 42, 95100 Selb, Germany

**Abstract**

It is well known that the concentration regimes in polymer and polyelectrolyte solutions can be identified by scaling laws for the relation between specific zero-shear viscosity and concentration. Recently, we have shown that the same is true for the infinite-shear viscosity plateau. The shear-thinning range is usually accessed by focusing on the viscosity functions for the respective concentration regime. For salt-free aqueous xanthan solutions, we find power-law dependencies of the specific viscosity on concentration throughout the entire shear-rate range. We distinguish six different concentration regimes. Apart from those already known for the zero-shear viscosity of polyelectrolyte solutions, i.e. dilute, semidilute unentangled, semidilute entangled and neutral semidilute entangled, we identify a linear regime for low shear rates at high concentrations, where the solution gels and a regime at both, higher concentrations and higher shear rates. Within some regimes, the power-law exponents change smoothly with shear rate, particularly, when deviating from the zero-shear viscosity plateau before the power-law of the viscosity function. Some regimes merge as their power-law exponents approach each other.

The fact that the regimes extend smoothly from the zero-shear regime into finite shear rates, i.e. away from thermodynamic equilibrium, shows that indicators such as critical concentrations remain valid at finite shear rates. This motivates us to interpret the data in the light of existing scaling laws and current knowledge about shear-rate dependent interaction mechanisms in polyelectrolyte solutions, particularly in xanthan solutions. It allows to follow the shift of relevant interaction mechanisms with shear rate. We think that the consideration of scaling laws under shear can be particularly helpful for identifying, for instance, thresholds for shear-induced disentanglement or disaggregation.

*Corresponding author: Andreas Wierschem
ORCID iD: 0000-0001-7927-2065
Lehrstuhl für Strömungsmechanik
Friedrich-Alexander-Universität Erlangen-Nürnberg
Cauerstr. 4, D-91058 Erlangen
Phone: +49-9131-85-29566
E-mail: andreas.wierschem@fau.de

# I. INTRODUCTION

The rheology of polymer and polyelectrolyte solutions has been studied extensively for many decades [1], [2]. At thermodynamic equilibrium, the concentration dependence of many solution properties such as polymer size or diffusion coefficient can be described by power laws. This also holds for relaxation times and zero-shear viscosity [3], [4]. Due to the concentration-dependent interaction between the polymers, the exponents are specific for the different concentration regimes, i.e. dilute, semidilute unentangled and semidilute entangled. They also depend on the solvent quality and differ between neutral polymers and polyelectrolytes [5], [6], [7]. A brief summary has been provided by Colby [3].

A lot of effort has been made to describe and understand the viscosity function of the polymer and polyelectrolyte solutions for the different concentration regimes [3], [8], [9], [10], [11], [12], [13], [14]. The solutions are commonly shear-thinning, with a power-law regime at intermediate shear rates and an infinite-shear viscosity plateau at high shear rates [15], [16]. They are usually well described by Carreau-Yasuda and Cross models [17], [18], [19], [20]. Particular emphasis has been dedicated to the power-law regime [21], [22], while the initial and terminal ranges of shear thinning have caught less attention. For the former, the main focus has been on the longest relaxation time [9], [23], [24]. result

Shear has a strong effect on the microstructure for complex soft materials, including polymer solutions [1], [2], [25], [26], [27]. Among others, it may result, for instance, in deformation, alignment, disentanglement and disaggregation of the constituents such as polymers. As a consequence, alignment or disentanglement may lead to reduced polymer-polymer interactions and higher concentrated solutions appear at high shear as if they were dilute [15], [16], [28], [29]. For salt-free xanthan solutions, we recently found that also the concentration-dependence of the infinite-shear viscosity can be described by power laws. More striking, the experimentally observed exponents are close to those for the dilute and semidilute unentangled regimes of polyelectrolyte solutions. The overlap concentration for the infinite-shear viscosity, however, is much higher than in the case of the zero-shear viscosity, and solutions entangled at zero shear show a concentration dependence typical for unentangled solutions in the infinite-shear viscosity plateau [15]. This might be surprising as the scaling laws initially derived for thermodynamic equilibrium appear far from it. Yet, one may argue that shear mainly deforms the polymers, changes their orientation, and promotes disentanglement [30]. As such, length scales in shear and flow direction may become different and critical concentrations may shift, leaving the major picture for the interactions between the polymers in the different concentration regimes intact. This in mind, one may wonder how the regimes transform from zero shear to infinite shear and whether the viscosity obeys concentration-dependent power laws in the shear thinning regime.

To study whether the concentration-dependence of the viscosity in the shear-thinning regime may be described with power-laws and how critical concentrations between the different regimes change with shear rate, we study salt-free aqueous xanthan solutions. As shown previously, its entire shear-rate range is accessible, i.e. from the zero-shear to the infinite-shear viscosity plateau [15]. Apart from that, aqueous xanthan solutions have been studied intensively [19], [20], [21], [22], [31], [32]. Xanthan is a polysaccharide produced by Xanthomonas campestris bacteria. The primary structure consists of a backbone similar to that of cellulose with charged trisaccharide side chains on alternating backbone residues [20], [33]. In its native state, xanthan forms a secondary structure in form of a 5-fold double helix stabilized by hydrogen bonds, where the double helical structure has a persistence length of about 120 nm [34]. Its solutions have a wide range of applications. They are used as viscosity modifiers, stabilizers, thickeners, drag reducers, drug delivery agents, flocculants, absorbents, dispersants in versatile industries such as coatings, adhesives, drilling fluids, foods, personal care, oil recovery, and agriculture among others [19], [21], [33], [35], [36].

The article is organized as follows: The solutions and the methods are described in Sec. II. The experiments and their results are reported and discussed in Sec. III. Finally, the conclusions are summarized in Sec. IV.

## II. MATERIALS AND METHODS

Aqueous xanthan solutions were prepared from xanthan powder (Sigma Aldrich) and deionized water (Sigma Aldrich). As stated by the supplier, approximately half of the terminal mannose residues of the xanthan are 4,6-pyruvated and most of the inner mannose residues are 6-acetylated. For samples from the same supplier, an acetate/pyruvate ratio of 6/5 was determined [37]. The molecular weight of xanthan was measured with size exclusion chromatography in water resulting in $M_w = 3.2 \pm 1$ MDa and $M_w/M_n = 1.2 \pm 0.2$ [15].

According to the supplier, the deionized water had a conductivity of less than 2 μS/cm. Water and xanthan were used as received. Solutions with xanthan concentrations up to 4 wt.% were prepared and observed to be optically clear. Since in the salt-free limit, studies at low polyelectrolyte concentrations in aqueous solutions are susceptible to impurities such as carbonic acid from the air and residual salts [38], we restricted our study to concentrations of no less than 0.0028 wt.%. The glassware and tools used for mixing and storage of the polymer solutions was carefully cleaned with either ethanol or acetone (reagent grade), then rinsed with purified deionized water to remove all traces of salt prior to use.

The solutions were prepared by dissolving the xanthan powder in the deionized water. They were mixed with a magnetic stirrer at room temperature continuously for 24 h or for 1-2 h before being allowed to rest for approximately 24 h. During that time, the sample was covered tightly with a Parafilm to minimize evaporation losses. After that, the solutions were centrifuged at 3000 rpm for 5 min to remove possible remaining bubbles before being stored into a refrigerator at 4°C for at least 1 h. Measurements were made 1-2 days after preparation. We have not observed any change in rheological properties of the solutions during that period nor any turbidity of the samples.

To study the broad viscosity range of the solutions at shear rates from $10^{-3}$ s$^{-1}$ to $1.5 \times 10^5$ s$^{-1}$, the rheological experiments were performed with different geometries. Low viscous samples up to concentrations of 0.28 wt.% were studied at shear rates below $10^2$ s$^{-1}$ with a concentric cylinder geometry on an MCR 702 Multidrive rotational rheometer (Anton Paar). The Searle system consisted of a C-CC20/TD (diameter: 22 mm) cup and B-CC20 (diameter: 20 mm, length 30 mm, cone tip 120°) bob. Temperature control was carried out using a CTD 180 measuring chamber. Filling quantity was 8.17 mL and distance between cone tip and beaker bottom was 4.23 mm (in accordance with DIN EN ISO 3219-2). Measurements were carried out after pre-shearing at a shear rate of 10 s$^{-1}$ for 30 s and 3 min of rest. End effects were taken into account by a correction factor of 1.1 in the conversion from torque to shear stress [39] and the data for the shear rate at the inner cylinder, $\dot{\gamma}_i$, was corrected by taking into account the change of the data points by the model-free correction [40]:

$$\dot{\gamma}_i = -\frac{\Omega}{\ln\beta}\left[1 - m\ln\beta - \frac{1}{3}(m\ln\beta)^2\right], \quad (1)$$

where $\Omega$ is the angular velocity and $\beta$ is the ratio of the inner to the outer cylinder. The slope correction, $m = d\ln\Omega / d\ln M$ with $M$ being the torque was obtained by using the central difference scheme.

In the same shear-rate range, higher viscous samples down to concentrations of 0.13 wt.% were studied with a cone-plate geometry with an opening angle of 1° and a diameter of 50 mm on an MCR 302 or alternatively on an MCR 301 rotational rheometer equipped with a P-PTD200 temperature control and a HPTD200 hood (Anton Paar). In this case, pre-shear was carried out at shear rates of 500 s$^{-1}$ or 1000 s$^{-1}$ for 30 s followed by 5 min of rest.

To access higher shear rates, we employed a home-made narrow-gap device based on an MCR 501 rotational rheometer (Anton Paar) [15], [16], [41], [42], [43], [44], [45], [46] with a parallel-disk geometry at a gap width of 20 µm. To arrive at this low gap width, the disks were replaced by glass plates (Edmund Optics). The lower plate with a diameter of 75 mm had an evenness of λ/4 (lower plate, 75 mm diameter), where λ is the testing wavelength (633 nm). As upper plate, we used glass plates of 50.8 mm diameter with an evenness of λ/4 or λ/10. The plates were aligned with actuators (Physik Instrumente). The gap width was set up and controlled with a confocal interferometric sensor (STIL). By these means, the uncertainty in the gap width was reduced to less than ±1 µm [41]. The experiments were carried out at variations in gap widths typically below ±0.7 µm. Temperature was maintained with a hood equipped with a temperature control system (H-Tronic TSM 125). For further details on the device and its alignment, we refer e.g. to Dakhil and Wierschem [41].

The narrow-gap device has a nominal gap volume of about 40 µL. Yet, we did not trim it and used intentionally large overloading. At the narrow gap used, this has only negligible impact on the data [47], [48]. For pre-shearing in the narrow-gap device, we followed the protocol established by Dakhil *et al.* [15]: Before pre-shear, the samples were tempered to 25°C. After an initial pre-shear at a shear rate of 1000 s$^{-1}$ for 30-60 s, the sample was sheared at 100 s$^{-1}$ for 30-60 s followed by 1-5 min of rest before starting the experimental runs. In order to make sure the sample is in a relaxed steady state, normal force and torque were monitored during pre-shear. Particularly at high shear rates, the normal force becomes significant. This can result in slight changes of the gap width. To account for this effect and to correct the shear rate and thus the viscosity, we monitored the gap width with the confocal interferometric sensor. The shear-stress data obtained with the parallel-disk configurations, $\tau_R$, was further corrected by taking into account the change of the data points according to [15], [49]:

$$\tau_R = \frac{2M}{\pi R^3}\left(\frac{3}{4} + \frac{1}{4}\frac{d\ln M}{d\ln\Omega}\right), \quad (2)$$

where $R$ is the plate radius. The slope correction, $d\ln M / d\ln\Omega$, was obtained by fitting the logarithmic data with a second- or third-order polynomial.

The temperature was fixed at 25.0°C. In the narrow-gap device, it varied by approximately ±0.5°C. Without having been heated beyond this temperature, xanthan is in the rather stiff ordered structure [37], which takes the form of a helical dimer [50]. Evaporation was minimized by placing wet tissue inside the measuring chambers. At shear rates below 1 s$^{-1}$, measurement time per data point corresponded to at least the inverse shear rate. At higher shear rates, it was 20 s.

## III. RESULTS AND DISCUSSION

We provide our data as a function of the dimensional shear rate. Although the relaxation time, particularly of diluted solutions, is a natural time scale to non-dimensionalize the shear rate, we do this for the following reasons: For semidilute unentangled solutions, the relaxation time is concentration dependent, while it is independent from concentration in the dilute regime [3]. There, however, it depends on the number of Kuhn monomers and their relaxation time, which depends not only on the molecular weight but also on the precise xanthan composition. As we have not studied dilute solutions, we refrain from scaling the shear rate by a relaxation time to avoid possible misunderstanding or mismatch. Therefore, we discuss the impact of shear on the scaling of the specific viscosity in terms of the dimensional shear rate.

FIG. 1 shows examples for the specific viscosity functions of xanthan solutions from different concentration regimes resulting from experimental runs with the different geometries. To compare the viscosities at equal shear rates, we interpolate our data using Akima splines. This interpolated data is indicated in FIG. 1 by open symbols. The data obtained with different geometries overlap well within the common shear-rate range. This also indicates that wall slip is not relevant in the considered shear rate region. Average deviations for overlapping data are usually less than 10%. For all concentrations, the viscosity levels off in the infinite-shear plateau at shear rates of about $10^4$ $s^{-1}$ or higher. Deviations from the zero-shear viscosity plateau occur beyond 0.1 $s^{-1}$ for lower concentrations while the zero-shear plateau diminishes for larger concentrations until a non-zero yield stress is observed.

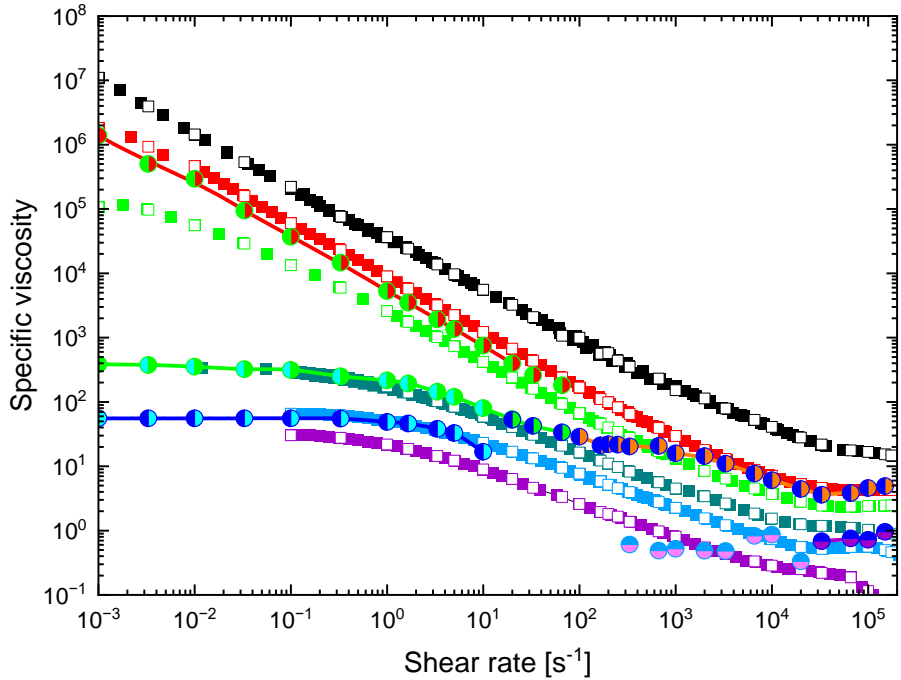

**FIG. 1.** Specific viscosity functions of xanthan solutions from different concentration regimes. Black, red, green, turquoise, blue, and purple squares show concentrations of 4.0 wt.%, 1.0 wt.%, 0.46 wt.%, 0.13wt.%, 0.046 wt.% and 0.013 wt.%, respectively. Open squares indicate interpolated data points. 2-color solid circles indicate the crossover between the different power-law regimes at the respective shear rates, see FIG. 3. Crossover values for missing data at low concentrations and shear rates were determined from the plateau values at the lowest shear rates measured.

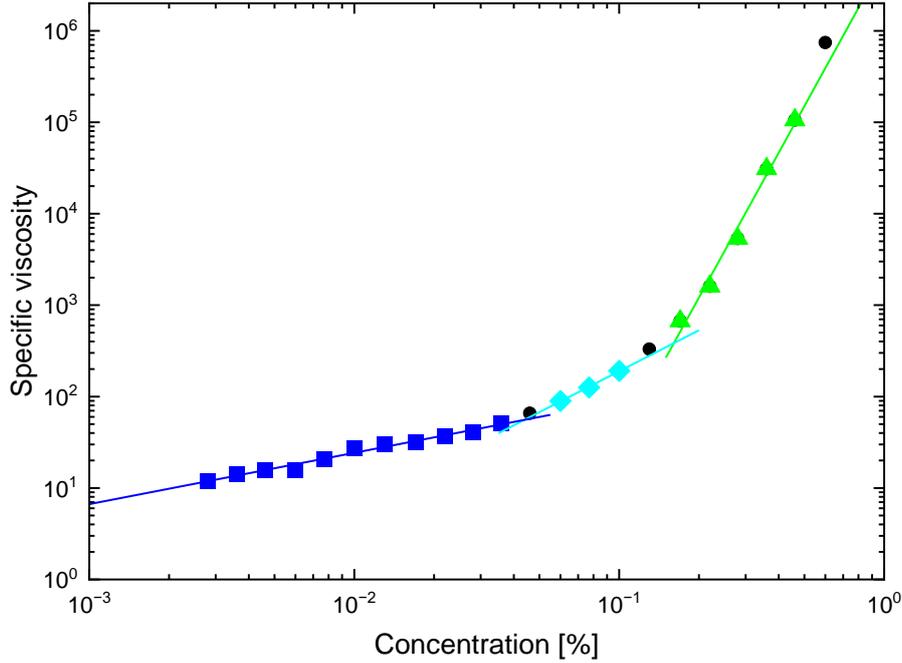

**FIG. 2.** Specific viscosity at zero shear as a function of xanthan concentration. Blue squares, turquoise diamonds, and green triangles indicate the semidilute unentangled and semidilute entangled polyelectrolyte regimes, and the neutral semidilute entangled regime, respectively. The data corresponds to shear rates of $10^{-3}$ s$^{-1}$ for concentrations larger than 0.22 wt.%, to shear rates of $10^{-1}$ s$^{-1}$ for concentrations lower than 0.13 wt.% and to shear rates of $10^{-2}$ s$^{-1}$ for the concentrations in between. Lines are power-law fits to the data of same color. Dark circles have not been fitted.

FIG. 2 shows the specific zero-shear viscosity of the xanthan solutions studied. We determined it from the plateau values at the lowest shear rates measured with a solvent viscosity of 0.889 mPas at 25°C [51]. For the highest concentration, we could not identify a zero-shear plateau, see FIG. 1. In this case, we used the viscosity value at the shear rate of $10^{-3}$ s$^{-1}$ as reference. The lines are power-law fits to the data of same color. For the lower concentrations, we arrive at an exponent of about 0.56±0.03, the narrow intermediate range has an exponent of about 1.5±0.1, while it is about 5.3±0.8 at the higher concentrations. The two former exponents are close to the theoretical values expected for semidilute unentangled and entangled polyelectrolyte solutions, i.e. 0.5 and 1.5 [3]. Here, we remark that for the semidilute unentangled regime, our exponent is somewhat larger than the theoretical value. Recently, an even larger exponent, i.e. about 0.68, has been reported, which has been related to higher chain flexibility [52]. The latter exponent is considerably larger than 3.9, the exponent for semidilute entangled solutions of neutral polymers in good solvents [3], which occurs when the correlation length, i.e. the mesh size of the semidilute solution, becomes smaller than the electrostatic blob size and electrostatic interactions become unimportant [4], [53], [54]. Instead, within fitting uncertainty the exponent is closer to 14/3, the exponent for a semidilute entangled solutions of neutral polymers in a θ solvent [3]. Yet, we suppose that the high exponent occurs due to the proximity to the gelling transition, where the zero-shear viscosity tends to infinity. Disregarding concentrations beyond 0.28 wt.%, for instance, reduces the exponent to 4.1±0.8. Furthermore, a regime with an exponent beyond 5 has been reported recently by Missi *et al.* for xanthan solutions at high concentrations before gelling sets in [50]. As will be discussed later, the exponent decreases continuously in this regime as the shear rate is raised. At shear rates of about 0.033 s$^{-1}$, the exponent is already comparable to the one expected for semidilute entangled solutions in good solvents.

The entanglement concentration, $c_e$, that results from the crossover of the fits is at about 0.044 wt.%, while the one for overlapping electrostatic blobs, $c_D$, is around 0.16 wt.%, i.e., the semidilute entangled polyelectrolyte regime is only encountered in a rather narrow range of concentrations. It is expected to occur only at sufficiently high molecular weight [54]. The specific viscosity at these concentrations is about 55 at entanglement concentration and about 380 at $c_D$. Hence, the specific viscosity at entanglement concentration is in the expected range of about 50 [55]. In the range of our study, the semidilute unentangled regime covers a decimal power in concentration. The overlap concentration, $c^*$, below which the solution is considered dilute, is usually at a specific viscosity of about 1 [3], which is well below the specific viscosity of the lowest concentrations studied. Prolonging the fit for our lower concentrations to a specific viscosity of 1 results in a concentration of about $3.5 \times 10^{-5}$ wt.%. Covering a concentration range of about 3 decades is expected for high molecular salt-free polyelectrolyte solutions [3].

The critical concentrations are close to those found by Wyatt and Liberatore for their salt-free xanthan solutions: They arrived at $c_e \sim 0.04$ wt.% for the entanglement concentration and at $c_D \sim 0.2$ wt.% [19]. Beyond $c_D$, also absolute viscosity values are quite similar. Below $c_D$, where electrostatic interactions are relevant, our zero-shear viscosities are considerably lower than those detected by Wyatt and Liberatore. As fermentation conditions influence molecular weight and the extent of pyruvic acid and acetal substitutions [21], [37], [56], [57], we suppose that this may have caused the difference in absolute values of the zero-shear viscosity.

The different concentration regimes for the zero-shear viscosity expand smoothly into the shear-rate dependent range. FIG. 3 provides examples for some shear rates. The straight lines indicate power-law fits to the data with coefficients of determination usually better than 0.98 except for the lowest concentrations at highest shear rates. As the diagram shows, it is possible to fit the entire shear-rate range with power laws. FIG. 4(a) shows the scaling exponents for the respective regimes. As appears from FIG. 3, we may distinguish six different regimes. The first three expand from the concentration regimes for the zero-shear viscosity, i.e. the semidilute unentangled polyelectrolyte regime (I), the semidilute entangled polyelectrolyte solutions (II), and the semidilute entangled solutions of neutral polymers (III). The fourth regime (IV) holds for the yields-stress materials at low shear rates. At higher shear rates, the regimes I and II merge into a regime that has been previously shown to have an exponent of about the value for the Fuoss law for the infinite-shear viscosity [15]. The two regimes III and IV also merge at higher shear rates. As the exponent differs from those found for polyelectrolyte solutions at zero shear, we call this a new regime (V). Finally, although we could not study the zero-shear viscosity of dilute solutions, the dilute regime appears at high shear rates (VI). In what follows, we will discuss our findings for each regime separately.

We start with the semidilute unentangled regime (I) as it covers the largest concentration range for the zero-shear viscosity. As the shear rate increases, we detect a slightly increasing exponent. From the initial value of about 0.56 for zero shear, it increases continuously to a maximum value of about 0.84±0.03 at a shear rate of about 100 $s^{-1}$, see FIG. 4(a). From here onward, the exponent decreases again smoothly to values of about 0.58 at shear rates of $10^5$ $s^{-1}$, where the infinite-shear viscosity is reached.

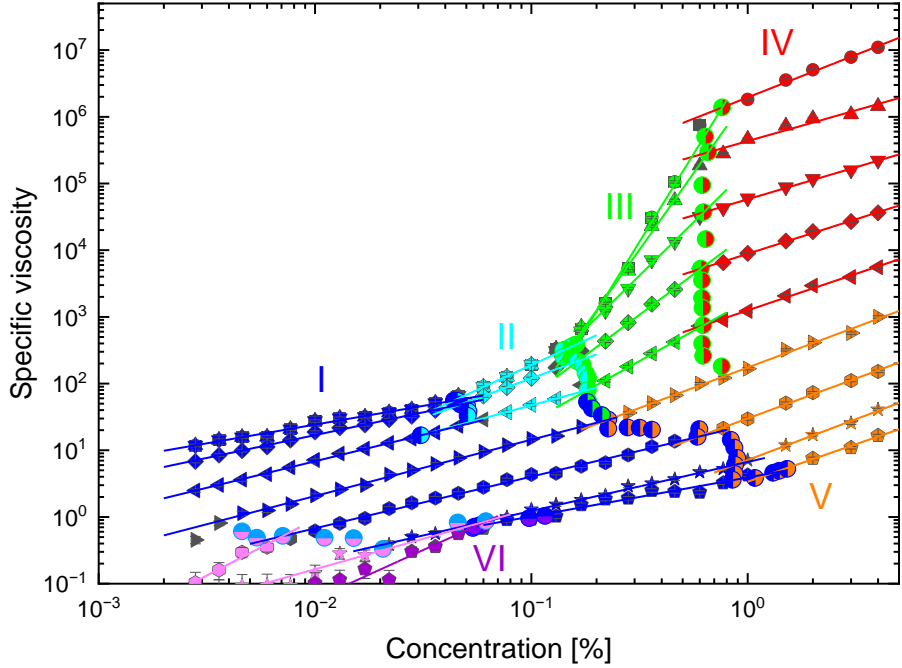

**FIG. 3.** Shear-rate dependent specific viscosity of the interpolated data as a function of xanthan concentration. ■, ●, ▲, ▼, ◆, ◄, ►, ●, ★, ⬠ indicate zero viscosity and shear rates of $10^{-3}$ s$^{-1}$, $10^{-2}$ s$^{-1}$, $10^{-1}$ s$^{-1}$, 1 s$^{-1}$, 10 s$^{-1}$, $10^{2}$ s$^{-1}$, $10^{3}$ s$^{-1}$, $10^{4}$ s$^{-1}$, $10^{5}$ s$^{-1}$, respectively. The Latin numbers enumerate the different scaling regimes, indicated by blue, turquoise, green, red, orange, and purple/pink symbols, respectively. Grey symbols had not been taken into account for the fits. Straight, solid lines are power-law fits to the data of same color. Bicolor circles indicate the critical concentrations between the regimes for the respective shear rates.

It is near at hand that the change in exponent with shear rate is related to the impact of deformation and orientation of the polyelectrolytes. Yet, due to electrostatic repulsion, polyelectrolytes in salt-free solutions are already extended at zero shear [55], [58], [59]. However, shear also provokes a redistribution of counterions: Changes in conformation affect the degree of ionization [7] and the average counterion-chain distance gradually increases with low shear rates resulting in an increase of the effective polyelectrolyte charge with shear [60], [61], [62]. These authors argue that nor chain conformation and orientation nor hydrodynamic drag could cause the displacement but it is rather a response to the changing electric-field distribution. During bypassing of neighboring chains their electric fields overlap and counterions explore more space as they experience additional electrostatic attraction from another chain. Hence, we expect this effect to be particularly relevant in the semidilute unentangled regime.

Studying sodium polystyrene sulfonate, Zheng *et al.* found that the local pH values increase with shear until saturating at a shear rate of about 5 s$^{-1}$, where the counterions are quite homogeneously distributed and the electric potential is supposed to having reached too high values for the counterions to overcome [60]. This shear-rate range coincides quite well with the range at which the exponent increases most significantly in our case, i.e. up to about 0.8 at a shear rate of 10 s$^{-1}$, see FIG. 4(a). A closer look at FIG. 1 reveals that it also coincides with the onset of shear-thinning before the power-law range is reached. Thus, the counterion redistribution has apparently a strong effect on the onset of shear-thinning. The higher effective polyelectrolyte charge may result in a concentration-dependent effective repulsion between the chains, which yields a higher flow resistance as concentration increases and hence a higher exponent. As the counterion-redistribution saturates, further alignment of the chains in flow direction may increase the effective distance between the chains at higher shear rates and thus mitigate the impact of the counterion redistribution.

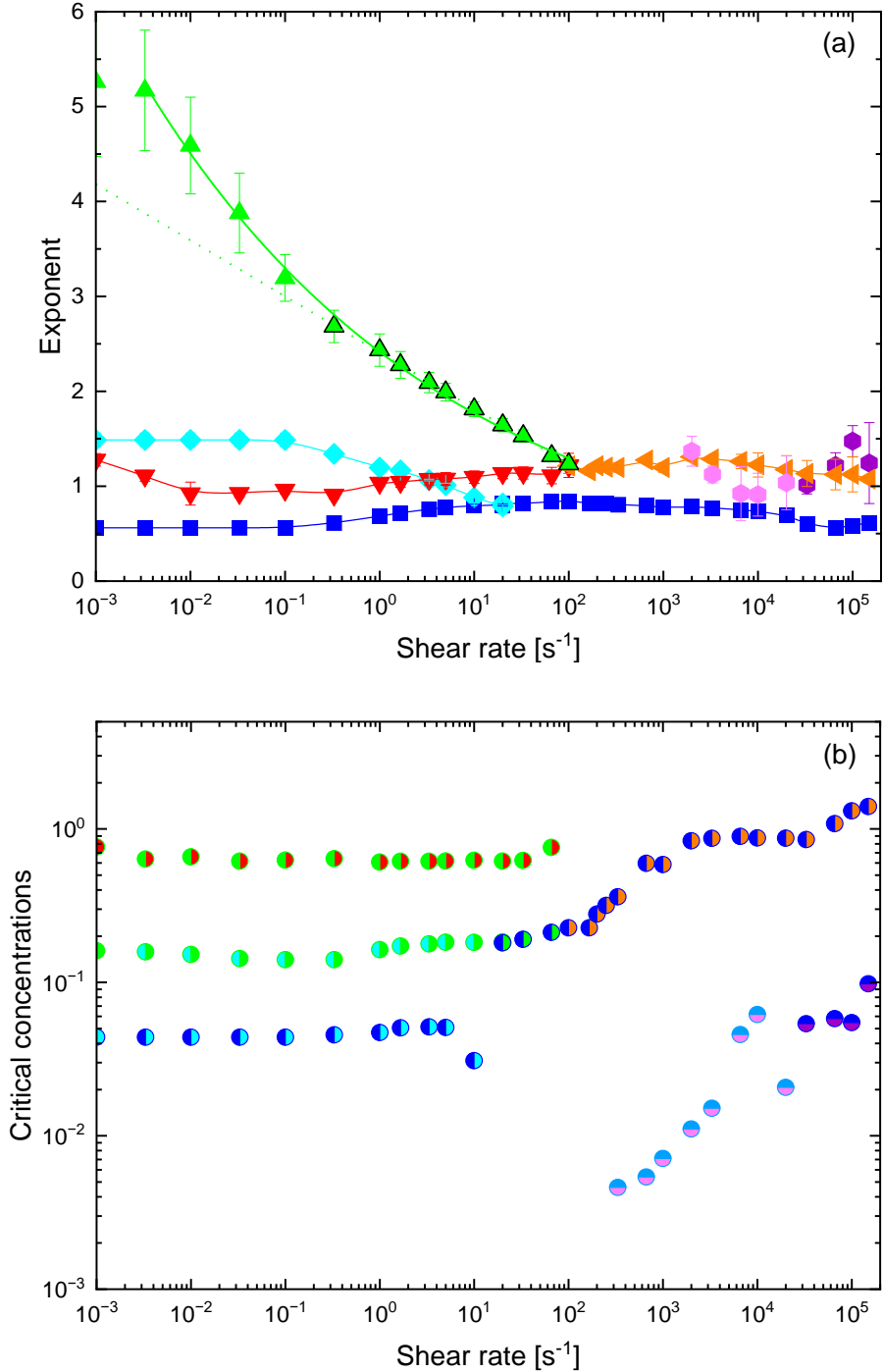

**FIG. 4.** Shear-rate dependent scaling exponents for the different concentration regimes (a) and critical concentrations (b). In (a), ■, ◆, ▲, ▼, ◀, ⬢ indicate the regimes I-VI from Fig. 3, respectively. The green solid curve in (a) is a power-law fit to the data of regime III in the shear-rate range covered by the curve; the dotted green line is a logarithmic fit to the exponent of regime III at higher shear rates, indicated by black-framed triangles. The bicolor circles in (b) indicate the crossover between the regimes as in Fig. 3. Crossover values for missing data at low concentration and shear rates were determined from the plateau values at the lowest shear rates measured.

The semidilute entangled polyelectrolyte regime (II) only extends to shear rates of about 10 s$^{-1}$, from where it merges with the semidilute unentangled polyelectrolyte regime (I), see FIG. 3 and FIG. 4. Like in the unentangled regime (I), this shear-rate range covers the onset of shear-thinning before the power-law regime, see FIG. 1. In the shear-thinning range from shear rates of about $10^{-1}$ s$^{-1}$ to the merger with regime II, the exponent decreases steadily to the value of the unentangled regime (I). It has been shown that in salt-free polyelectrolyte solutions, dipoles form when counterions adsorb on the charged chains, resulting in an attractive force between the chains that can promote aggregation at higher polyelectrolyte concentrations [7]. It is not clear to us whether the polyelectrolyte entanglement results from this dipole interaction. Yet, with shear indications of a loss of entanglement are reported [63]. Hence, we would expect that direct interaction between polyelectrolytes are smoothed as the chains start to align under shear, resulting in an ever less impact of concentration on entanglement and hence a lower exponent for the dependence on concentration as the shear rate increases. Finally, at the shear rate, where the regime II merges with regime I, the solution appears to be completely disentangled and behaves like an unentangled one with a considerable redistribution of counterions.

While the exponents in the two former regimes do not change by a factor of 2 or more, in the third regime, i.e. for neutral semidilute entangled solutions (III), the exponent decreases continuously to about 1.2 before it merges with regime IV. As indicates the solid green curve in FIG. 4(a), the decline can be described well by a simple power-law fit to the data with an exponent of $-0.135 \pm 0.002$. FIG. 4(a) also shows that focusing on shear rates beyond 0.1 s$^{-1}$, where the fitting uncertainty is much smaller, the data can be described by a logarithmic fit. In this case, we arrive at shear rates of $10^{-3}$ s$^{-1}$ at an exponent of about 4.2, which is close to that for semidilute entangled solutions in good solvents [3].

As shows FIG. 1, the decline of the exponent in the third regime happens in the shear-thinning power-law region as well as in the range where the viscosity functions particularly of lower concentrations are now yet in the power-law regime. It is particularly remarkable that the shear-rate dependent specific viscosity can be described by power-laws in this regime as degree of entanglement. In the entangled regime, the average number of entanglements increases with concentration at zero viscosity. In the shear-thinning range, chain alignment apparently has a progressive impact on the number of entanglements or on the relation between effective mesh size and concentration, resulting in the decreasing exponent with shear in a similar fashion as discussed before for the polyelectrolyte entangled regime (II).

The final exponent in regime III, before merging with regime IV, is close to those for neutral semidilute unentangled solutions in good solvents, i.e. 1.3, as well as for semidilute entangled polyelectrolyte solutions, i.e. 1.5 [3]. As we have seen before, there is a lower limit for the specific viscosity to switch from neutral entangled to polyelectrolyte entangled at low shear rates, see FIG. 3. Analogously to the entangled polyelectrolyte regime II, one may expect that the number of entanglements reduces continuously with shear rate until the solution becomes finally completely unentangled. Hence, it appears near at hand that the solution is neutral and unentangled when the regime III merges with regime IV. This interpretation is supported by the observation that the crossover concentration between regime II and regime III hardly changes before regimes I and II merge, see FIG. 4(b). Thus, the polyelectrolyte concentration is apparently high enough to behave as a (semidilute) solution of neutral polymers. At the shear rate where the regimes I and II merge, the specific viscosity at the concentration of the crossover between the merged regime and regime III has dropped to about 50, a value at which the entanglement concentration is expected [55]. As the shear rate increases the specific viscosity at crossover decreases further.

At still higher concentrations, the solutions gel, hence, have no zero-shear viscosity. In xanthan solutions, weak gels form due to the formation of aggregates through entanglement, hydrogen bonding in intermolecular association due to acetate residues or hydrophobic interactions between pyruvate substituents, giving rise to a weak three-dimensional network [22], [31], [64]. In this concentration range, a high supramolecular organization has been reported at room temperature [65]. A dispersed anisotropic phase has been reported, which volume fraction increases with concentration and which shows characteristics of liquid crystals [64].

With shear, the intermolecular structures break up and smaller structures survive [66]. The onset of gelling comes along at small shear rates with a change from regime III to regime IV. As shows FIG. 4(a), their specific viscosities in regime IV increase about linearly with concentration. The linear relation between specific viscosity and concentration supports the former statement on a break-up of the network into aggregates that interact with each other during shear. With increasing shear rate, the size of the aggregates is supposed to decline until they disappear. FIG. 3 and FIG. 4(b) show that the critical concentration for the onset of this regime appears to be hardly affected by shear rate if at all.

Before regime IV merges with regime III in the shear-rate range between about 66 $s^{-1}$ and 100 $s^{-1}$, the exponent in regime IV slightly increases, see FIG. 4(a), until both regimes become indistinguishable. The slight increment may be due to the increasing relevance of the interactions that govern regime III as the aggregates become smaller. FIG. 3 shows that the new regime V extends into the higher shear-rate range until the infinite-shear viscosity plateau is reached. Hence, apart from the diluted and semidiluted unentangled polyelectrolyte regimes identified for the infinite-shear viscosity plateau in a previous study up to concentrations of 1 wt.% [15], we find another one at higher concentrations.

In regime V, the exponent remains about constant within a narrow band around 1.2. As previously mentioned, the value coincides well with that for the zero-shear viscosity of good neutral semidilute unentangled solutions [3]. As it occurs at the high-shear end of the semidilute entangled regime for neutral polymers in good solvents (regime III) and that of the concentrated system with aggregates (regime IV), it is near at hand that the chains in regime V are indeed unentangled and behave practically like neutral ones. FIG. 3 shows that the critical concentration for the transition from the merged regimes I and II, i.e. the semidilute unentangled polyelectrolyte regime, into regime V shifts towards higher values with increasing shear rate. The specific viscosity at critical concentration drops to around 4 at shear rates, where the viscosity functions levels off. The shift in critical concentration with shear rate and the decline of the critical concentration point towards the important role of chain alignment for the critical concentration for overlap of electrostatic blobs.

At last, we consider regime VI. The dilute region was not accessible for the zero-shear viscosity due to sample exposure to open air and thus to impurities. Yet, at low concentrations we find noticeable deviations from regime I at shear rates of the order of $10^2$ $s^{-1}$, from where it expands to higher concentrations with shear rate. At shear rates of the order of $10^3$ $s^{-1}$, the data follows roughly a linear trend. Yet, at the small values of the specific viscosity, data scatter is significant. Together with the fact that the exponent in regime I is still beyond 0.7, this results in significant scatter for the overlap concentration. In Figures 1, 3 and 4, this data is indicated by pale data points. In the shear-rate range of the order of $10^4$ $s^{-1}$, where the viscosity levels off also at higher concentrations, the exponent in regime I is smaller and both regimes are more easily distinguishable. The exponent in this region is about one, see FIG. 4(a). As expected for the zero-shear viscosity [3] and also found in a previous study for the infinite-shear viscosity plateau [15], this regime occurs at specific viscosities below one. Thus, the chains are so much aligned at these higher shear rates that they hardly interact with each other although they overlap at zero shear. Different from regime I, the chains hardly meet each other and thus the counterion distribution is supposed to remain mainly unaffected.

## IV. CONCLUSIONS

It is well known that the major interaction mechanisms in polymer and polyelectrolyte solutions change with concentration, leaving their fingerprints in specific scaling laws for the specific zero-shear viscosity on concentration. Recently, we have shown that the same is true for the infinite-shear viscosity plateau. While the shear-thinning range is usually accessed by focusing on the viscosity functions for the respective concentration regime, we show here that the shear-rate dependent solution viscosity of xanthan also obeys concentration-dependent power laws, which smoothly continue from the plateaus. For the present case of salt-free aqueous xanthan solutions, we could identify six different concentration regimes. Apart from those already known for the zero-shear viscosity of polyelectrolyte solutions, i.e. dilute, semidilute unentangled, semidilute entangled and neutral semidilute entangled, we found a regime for low shear rates at high concentrations where the solution gels and a regime at higher concentrations and shear rates. The transitions between the regimes extend from zero shear to shear thinning and finally to the infinite-shear viscosity plateau, showing that indicators such as critical specific viscosity at crossover and critical concentration for gelling remain valid at finite shear rates. In particular, the crossover concentrations between the different regimes at zero-shear viscosity remain practically unaffected by shear until the regimes merge.

We interpret the data in the light of existing scaling laws and current knowledge about shear-rate dependent interaction mechanisms in polyelectrolyte solutions, particularly in xanthan solutions. This allows to follow the shift of relevant interaction mechanisms with shear rate. We think that the consideration of scaling laws under shear can be particularly helpful for determining, for instance, parameter regimes for specific interaction mechanisms to become relevant and identifying the threshold for shear-induced disentanglement and disaggregation. In the light of the present data, it appears that dilute solutions occur at specific viscosities below one not only at zero-shear conditions but also at higher shear rates. The exponent for semidilute unentangled polyelectrolytes solutions is slightly affected by shear rates. The exponent of semidilute entangled polyelectrolyte solutions declines with shear rate until this regime merges into the unentangled one. Also, for the entangled neutral solution, which occurs at higher concentrations, the exponent deteriorates continuously until that for neutral semidilute unentangled solutions is reached. This is where it merges with the gelling regime. Hence, at higher shear rates, three regimes remain: A dilute one, a semidilute unentangled polyelectrolyte solution and a neutral semidilute unentangled solution.

**Acknowledgments**

We thank Mrs. D. Tosunoğlu for preliminary measurements.

# AUTHOR DECLARATIONS

## Conflict of Interest

The authors have no conflicts to disclose.

## Author Contributions

**Ammar El Menayyir:** Data curation (equal); investigation (lead); project administration; validation; writing/original draft preparation; writing/review & editing (equal). **Markus Neuner:** Data curation (equal); investigation (lead); project administration; validation; writing/review & editing (equal). **Polina Fuks:** Investigation (equal); writing/review & editing (equal). **Vahid Amirgouneh Zadeh Alashloo:** Investigation (equal); writing/review & editing (equal). **Halim Altuntas:** Investigation (equal); writing/review & editing (equal). **Zehau Luo:** Investigation (equal); writing/review & editing (equal). **Melike Özgül:** Investigation (equal); writing/review & editing (equal). **Claudia Seeberger:** Investigation (equal); writing/review & editing (equal). **Sharadwata Pan:** Writing/original draft preparation; writing/review & editing (equal). **Andreas Wierschem:** Conceptualization; data curation (lead); formal analysis; funding acquisition; methodology; project administration (lead); resources; supervision; validation (lead); visualization; writing/original draft preparation (lead); writing/review & editing (equal).

# DATA AVAILABILITY

The data that support the findings of this study are available from the corresponding author upon reasonable request.